\documentstyle[aps,multicol,epsfig]{revtex}
\begin{document}
\draft
\title{Quantised motion of an atom in a Gaussian-Laguerre beam}
\author{J. Twamley}
\address{Laser Optics and Theory Groups\\
Blackett Laboratory\\
Imperial College\\
London, SW7 2BZ, UK}
\date{\today}
\maketitle

\begin{abstract}
We quantise the centre of mass motion of a neutral Cs atom in the presence
of a classical Gaussian-Laguerre${}_{10}$ light field in the large detuning limit.
This light field possesses orbital angular momentum which is transferred to
the atom via spontaneous emissions. We use quantum trajectory and analytic
methods to solve the master equation for the 2d centre of mass motion with
recoil near the centre of the beam. For appropriate parameters, we
observe heating in both the cartesian and polar observables within a few
orbits of the atom in the beam. The angular momentum, $\hat{L}$, shows
a rapid diffusion which results in $\langle \hat{L}\rangle $ reaching a
maximum and then decreasing to zero. We compare this with analytic results
obtained for an atom illuminated by a superposition of Gaussian-Laguerre modes which possess no angular momentum, in the limit of no recoil.
\end{abstract}


\begin{multicols}{2}
%
\section{Introduction}

Recently there has been a growing interest, both in theory and experiment,
in the interaction between matter and light fields possessing orbital
angular momentum \cite{INTEREST}. The typical light fields used in these
works are the Gaussian-Laguerre modes (G-L). Experimental work has
concentrated on two areas; the study of the motion of microscopic particles
in Gaussian-Laguerre beams \cite{FRIESE}, and the use of such beams in atom trapping 
\cite{ERTMER_TRAP}. The theoretical work so far has solely concentrated on
the semi-classical description of the motion of matter in 
G-L beams \cite
{ALAN_INTRO}. As neutral atom cooling improves, a truly quantum description
becomes necessary. Indeed, recent experiments in atomic guiding using
evanescent waves within a hollow optical fibre verge on the quantum regime 
\cite{JAPAN}. In this paper we examine the quantised motion of a neutral
atom in a Gaussian-Laguerre${}_{10}$ mode, taking into account
the spontaneous recoil while in the far detuned limit. We are primarily
interested in the quantised transfer of the orbital angular momentum to the
centre-of-mass (COM) of the atom and how this appears in the quantal
evolution. In the limit where the incident light is far detuned from the
atom's natural frequency, we find that the transfer of angular momentum is
imparted through the dissipation suffered by the atom through spontaneous
recoil. We derive a master equation for the COM motion in the limit of large
detuning for motion near the centre of the beam which includes the momentum
kick from spontaneous recoil. We select the beam and atomic parameters which
give an atom that can be easily cooled to the lowest COM eigenstate of the
G-L beam and which suffers a large amount of spontaneous recoil. This results
in a rapid increase of angular momentum over the orbital time of the atom
around the center of the beam. The atom, because of the dissipation caused by the
spontaneous recoil, also heats up. This is indicated by an increase in the
variances of $x$ and $p$.

For the case of a neutral Cesium atom, we solve the master equation using
quantum trajectories and numerically compute expectation values and
variances for the atom's orbital angular momentum, cartesian position and
momentum. We solve for an atom initially in a minimum uncertainty state (1)
at the centre of the beam and, (2) offset from the centre. We find that the
variances increase with time as the system heats due to the diffusion caused
by the spontaneous recoil. Surprisingly, we also find both in (1) and (2),
that the diffusion in the angular momentum $\hat{L}$, is so rapid that $%
\langle \hat{L}\rangle $ reaches a maximum and then decreases to zero. This
is to be contrasted with the semiclassical work \cite{ALAN_INTRO}, which
shows a general increase in the atom's semiclassical angular momentum while
in the harmonic regime of the light field potential. We finally compare this system to a
similar master equation with no cross terms and no momenta transfer on
recoil. This corresponds to an atom coupled to a superposition of G-L${}_{+10}$ and G-L${}_{-10}$ modes in the semiclassical limit. Here no angular momentum is
transferred but the variances in both $x$ and $y$ remain much smaller than
in the case where recoil is included.

\section{Gaussian-Laguerre Modes and Master Equation}

Gaussian Laguerre modes can be produced through cylindrical lens mode
converters \cite{CYLLENS}, spiral phase plates \cite{SPIRAL} and through
computer generated holograms \cite{RODNEY}. The general form for the
electric field for the Gauss-Laguerre${}_{l0}$ mode is \cite{RODNEY} 
\begin{equation}
E_{l0}^{+}=E_{0}\left( \frac{r}{w}\right) ^{l}e^{-(r/w)^{2}}e^{\pm il%
\theta }e^{-i\frac{kr^{2}}{2R}}e^{i(kz+\Phi )}  \label{eq1}
\end{equation}
where $w$ is the beam waist, $R$ is the wavefront radius of curvature and $%
\Phi $ is the Guoy phase shift. For most applications the paraxial
approximation is quite good. In this approximation it has been shown \cite
{ENK,POWER_JULY} that the Guoy phase and Rayleigh curvature are very small
for Gaussian-Laguerre modes near the beam waist and we shall ignore them.
We thus restrict
the atom's motion to a two dimensional plane transverse to the beam
propagation
direction, positioned at the beam waist. We shall concentrate on the $l=+1$
mode. Near the centre of the beam the exponential dependence on $r$ is small
and we shall also ignore it.

When the system is near resonance the internal electron degrees of freedom 
are strongly coupled to the external spatial degrees of freedom of the atom. 
To solve this fully quantum mechanically, for two spatial degrees of freedom,
 is virtaully impossible. To simplify matters, we work in the far detuned
 regime. In this regime, the master equation, in the dipole approximation,
and adiabatically eliminating the dynamics of the internal excited state
(assuming a two-level atom), takes the form \cite{DYRTING} 
\begin{eqnarray}
\frac{d}{dt}\rho &=&-\frac{i}{\hbar }\left[ \frac{1}{2m}(p_{x}^{2}+p_{y}^{2})+%
\frac{\hbar |\hat{\Omega}|^{2}}{\Delta |\nu |^{2}},\,\rho \right]\\\nonumber
& +&\frac{%
\Gamma }{2\Delta ^{2}|\nu |^{2}}\left( 2{\cal N}\hat{\Omega}\rho \hat{\Omega}%
^{\dagger }-\left\{ \hat{\Omega}^{\dagger }\hat{\Omega},\,\rho \right\}
\right) \;\;,  \label{master1}
\end{eqnarray}
where $\hat{\Omega}$ is the spatially dependent Rabi frequency, $\Delta $ is
the detuning, and $|\nu |^{2}=1+\Gamma ^{2}/(2\Delta ^{2})$. The super
operator ${\cal N}$, describes the effect of a spontaneous emission on the
atom \cite{KAZENTSEV}, 
\begin{equation}
{\cal N}\rho =\int \,d\vec{n}\,\Phi (\vec{n})\,e^{ik\vec{n}\cdot \vec{x}%
}\rho e^{-ik\vec{n}\cdot \vec{x}}\;\;,
\end{equation}
where $\Phi (\vec{n})$ is the dipole emission probability function which is
defined to be \cite{RUSSIAN} 
\begin{equation}
\Phi (\vec{n})=\frac{3}{8\pi }\left( 1-\frac{(\vec{d}\cdot \vec{n})^{2}}{%
\vec{d}\cdot \vec{d}}\right) \;\;,
\end{equation}
for plane-polarised incident light and 
\begin{equation}
\Phi (\vec{n})=\frac{3}{16\pi }\left( 1+(\vec{z}\cdot \vec{n})^{2}\right)
\;\;,  \label{circular}
\end{equation}
for circularly polarised incident light. Here $\vec{n}$ is the unit vector
giving the direction of the spontaneously emitted photon, $\vec{d}$ is the
direction of the atomic dipole, and $\vec{z}$ is along the direction of beam
propagation.

Before continuing we will discuss the validity of the two assumptions made
above. From considerations that will follow, we will choose to work with the 
$P_{3/2}\rightarrow S_{1/2}$ transition in Cs. Firstly, to consider this to
be a two-level atom we must excite the $P_{+3/2}\leftrightarrow S_{+1/2}$ or 
$P_{-3/2}\leftrightarrow S_{-1/2}$ transitions. This can be done with
incident $\sigma _{\pm }$ circularly polarised light. Plane polarised
incident light does not result in a closed 2-level atomic system. Thus, we
must take (\ref{circular}) as the angular probability distribution for the
direction of spontaneous emission of photons.

Secondly, we have chosen to work in the dipole approximation.  We now explain
why this approximation is still valid even though the COM is quite cold.
The electronic
transitions occur on much faster timescales than the dynamics of the COM.
Thus, although the COM is cooled to where the de-Broglie wavelength of the COM degree
of freedom is relatively large, the electronic degrees of freedom with which
the atom and light interact are not cooled and can rapidly follow the
slow COM motion. Thus the electric dipole
approximation holds when the spatial gradients of $|\Omega |^{2}$ are small
on the scale of the electronic clouds involved in the transition and
not on the scale of the de-Broglie wavelength of the COM.
To obtain
field gradients large enough to significantly excite higher multipoles one
must be working in the extreme non-paraxial regime. The degree to which the
Gauss-Laguerre modes are non-paraxial and thus couple to higher multipoles
is extremely small. For the physical model discussed in this paper we can
estimate the ratio of the azimuthal spin-orbit force to the transverse electric
field force to be $F_{s-1}/F_{\phi}\sim 6\times 10^{-6}$, \cite{ALLEN_MAY_1996}.
The omission of higher multipoles is therefore a very good approximation
for these modes. One consequence of this approximation is that the exchange
of spin angular momentum is conserved separately from the exchange of
orbital angular momentum \cite{VANENK_1994}, i.e. there is very little
spin-orbital coupling. Since $\Phi (\vec{n})$ is insensitive to the sign of
the helicity of the incident light and the Clebsch-Gordan coefficients for
the transitions in Cs are symmetric under $\sigma \rightarrow -\sigma $ we
are forced to conclude that the quantum evolution of the COM is highly
insensitive to the helicity of the incident light. This conclusion, although
quite reasonable on the atomic scale, seems to be at odds with recent
experiments on the ``classical'' motion of micro-spheres illuminated by
circularly polarised Gaussian-Laguerre beams \cite{FRIESE,INTEREST}.
There are two possible physical mechanisms which could generate the
observed rotation in mesoscopic particles. The transfer of spin angular
momentum to orbital angular momentum could be mediated through some
bifrefringence present in the material, akin to Beth's original experiment.
However, the transfer could also
be mediated through non-radiative off-resonant couplings, i.e.
absorption.
Although the first mechanism might play some small part in the experiments,
recent results seem to indicate that absorption is the primary mechanism
which couples both the spin and orbital angular momentum of the light, to
the orbital angular momentum of the mesoscopic particles
\cite{NEWINTEREST,FRIESE}.
However, in the quantum system studied here, the spin-orbit coupling, 
as we have argued above,
for a single atom interacting with a Gaussian-Laguerre beam should be
quite small.
An experiment to probe the
dependence of the COM motion of ultra-cold atoms on the incident light's
helicity would be quite illuminating.

\subsection{Master Equation}

We begin by recasting the master equation (\ref{master1}) into a more
convenient form. We first set $\hat{\Omega}=(\Omega _{0}/w)\,\tilde{\Omega}$
where $\tilde{\Omega}=\hat{r}\exp \,(i\hat{\theta})=\hat{x}+i\hat{y}$. Next,
we obtain the orbital frequency of the atom in the GL${}_{10}$ mode in the
harmonic approximation to be 
\begin{equation}
\omega _{s}^{2}=\frac{2\hbar \Omega _{0}^{2}}{m\Delta |\nu |^{2}w^{2}}\;\;,
\end{equation}
where $m$ is the mass of the atom. We further include the following
rescalings to give a master equation with coefficients of order unity
in the 
dimensionless quantities, $\bar{X}, \bar{Y}, \bar{P}_{\bar{x}}$, and $\bar{P}_{\bar{y}}$, 
\begin{eqnarray}
\hat{x}=\alpha _{x}\bar{X}\;\;,\;\; &&\hat{p}_{x}=\alpha _{p}\bar{P}_{\bar{x}%
}\;\;, \\
\hat{y}=\alpha _{x}\bar{Y}\;\;,\;\; &&\hat{p}_{y}=\alpha _{p}\bar{P}_{\bar{y}%
}\;\;, \\
\alpha _{x}=\frac{1}{\sqrt{\beta }}\sqrt{\frac{\hbar }{m\omega _{s}}\;\;,\;\;%
} &&\alpha _{p}=\frac{1}{\sqrt{\beta }}\sqrt{\hbar m\omega _{s}}\;\;,
\end{eqnarray}
with $\tau =\omega _{s}t$. This gives $[\bar{X},\bar{P}_{\bar{X}}]=i\beta $,
where $\beta $ serves as the rescaled Planck's constant. Letting $\eta
=\Gamma /(4\Delta \beta )$, we finally obtain 
\begin{eqnarray}
\frac{d}{d\tau }\rho &=&-\frac{i}{2\beta }\left[ \bar{P}_{\bar{x}}^{2}+\bar{P}%
_{\bar{y}}^{2}+\bar{X}^{2}+\bar{Y}^{2},\,\rho \right]\\\nonumber
& -&\eta \left( 2{\cal N}(%
\bar{X}+i\bar{Y})\rho (\bar{X}-i\bar{Y})-\left\{ \bar{X}^{2}+\bar{Y}%
^{2},\,\rho \right\} \right) \;\;.  \label{master2}
\end{eqnarray}

For an atom cooled to the recoil limit, $P_{recoil}=\hbar k$ or 
\begin{equation}
\bar{P}_{r}=\sqrt{\beta }\sqrt{\frac{\hbar k^{2}}{m\omega _{s}}}\;\;.
\end{equation}
If the atom is initially in a pure, minimum uncertainty state, we have 
\begin{equation}
\Delta \bar{P}_{r}=\sqrt{\beta }\sqrt{\frac{\hbar k^{2}}{m\omega _{s}}}%
\;\;,\;\;\Delta \bar{X}_{r}=\frac{\beta }{2\Delta \bar{P}_{r}}\;\;,
\label{uncertainty}
\end{equation}
while the effect of the exponential in the superoperator ${\cal N}$ is to
shift the momentum, 
\begin{equation}
e^{-ik\alpha _{x}\bar{X}}\,\bar{P}_{\bar{X}}\,e^{+ik\alpha _{x}\bar{X}}=\bar{%
P}_{\bar{X}}+\mu \beta \;\;,\label{equ13}
\end{equation}
where $\mu =k\alpha _{x}$. We note that the master equation (\ref{master2})
is more involved than the majority of master equations previously discussed
within the literature. The complications here are twofold. In the Linblad
form for the master equation (\ref{master2}), the output channel is not just
a momentum jump (as it is in the near-resonance situation) but is a
combination of a diffusion in space and momenta. Secondly, we note the
presence of cross terms between $\bar{X}$ and $\bar{Y}$ in the dissipation.
This cross-coupling frustrates the derivation of any simple equation or set
of equations for the time evolution of ensemble averages.
Finally, if $\mu\beta$ is much smaller than the momentum variance of
the wavepacket $\langle\Delta\bar{P}\rangle$, then the exponential
operator in (\ref{equ13}) can be expanded to first order in $\bar{X}$.
One can then explicitly perform the stochastic average over $d\vec{n}$ 
to arrive at a Linblad type
master equation. This can be converted into a c-number Fokker-Planck
equation which (since it is at most quadratic), will give, in the large
energy regime, expectation values which agree with the semi-classical
theory. However, below we will choose the system parameters such that
$\mu\beta$ is large and thus one must solve the full quantum dynamics
of (\ref{master2}).

\subsection{Quantum Trajectories}

The master equation (\ref{master2}) must be solved numerically. We will use
the method of Quantum Trajectories \cite{QT}. However, we will be able to
solve for the non-unitary evolution analytically between jumps. This is
possible only in the harmonic approximation for motion near the centre of
the beam. To use this technique we rewrite the master equation (\ref{master2}%
) in the form 
\begin{eqnarray}
\frac{d}{d\tau }\rho &=&-\frac{i}{\beta }\left[ H_{0},\rho \right]\\\nonumber
 &+&\int
\,d^{2}\vec{n}\,\left[ 2C(\vec{n})\rho C^{\dagger }(\vec{n})-\left\{
C^{\dagger }(\vec{n})C(\vec{n}),\rho \right\} \right] \;\;,
\end{eqnarray}
where 
\begin{eqnarray}
C(\vec{n}) &\equiv &\sqrt{\eta \Phi (\vec{n})}(\bar{X}+i\bar{Y})\,e^{i\mu
(\epsilon _{x}\bar{X}+\epsilon _{y}\bar{Y})}\;\;,  \label{JUMPOP} \\
2H_{0} &\equiv &\bar{P}_{\bar{X}}^{2}+\bar{P}_{\bar{Y}}^{2}+\bar{X}^{2}+\bar{%
Y}^{2}\;\;,
\end{eqnarray}
and where $\epsilon _{x,y}$ are the $x$ and $y$ components of the recoil
direction vector $\vec{n}$. To apply the method we first choose an initial
state $\rho _{0}=\sum_{i}|\Psi _{i}\rangle \langle \Psi _{i}|$. In our case
we will set the initial state to be a pure coherent state. From the ensemble 
$\rho _{0}$ we choose a particular $|\Psi _{i}\rangle $ at random and apply
the following procedure. We generate a uniform random variable $\zeta \in
[0,1]$, and evolve the pure state $|\Psi _{i}\rangle $ using the non-unitary
Hamiltonian $H_{non}$, 
\begin{equation}
H_{non}\equiv H_{0}-i\beta \int \,d^{2}\vec{n}\,C^{\dagger }(\vec{n})C(\vec{n%
})=H_{0}-\delta (\bar{X}^{2}+\bar{Y}^{2})\;\;,
\end{equation}
where $\delta =i\beta \eta $. We evolve for a time $\tau _{\zeta }$ such
that, $|\langle \Psi (\tau _{\zeta })|\Psi (\tau _{\zeta })\rangle
|^{2}=\zeta $. We then apply the ``Jump'' operator $C(\vec{n})$, to $|\Psi
(\tau _{\zeta })\rangle $, $C|\Psi \rangle \rightarrow |\Psi ^{\prime
}\rangle $, where the vector $\vec{n}$ is generated by two random numbers
taken from the distribution $\Phi (\vec{z},\vec{n})$, (\ref{circular}). We
then renormalise the state $|\Psi ^{\prime }\rangle $ and repeat the
procedure beginning with the generation of a new uniform random waiting
time, $\zeta $. At set intervals $\tau _{s}$, we store the state. Repeating
this whole procedure, starting with randomly sampled $|\Psi _{i}\rangle $
from the ensemble $\rho _{0}$, we generate estimates for the density
matrices at the times $\tau _{s}$. From these estimates we can calculate
expectations and variances for various observables. In practice, unless the
final quantities involve the computation of off-diagonal elements of the $%
\rho (\tau _{s})$, one only needs to store the incremental values for these
quantities and not the complete $\rho (\tau _{s})$ themselves. This produces
enormous savings on hard memory usage since the storage size of a single
state $|\Psi _{i}\rangle ,$ in a truncated Fock state basis of 40, in two
dimensions, is approximately 25KB in double precision.

As stated above, in the harmonic approximation, $H_{non}$ is quadratic and
the propagator, $U_{non}=\exp \,(iH_{non}\tau )$, can be evaluated
analytically in the Fock state basis. This basis is constructed through the
operators 
\begin{equation}
a_{\bar{X}}^{\dagger }=(\bar{X}-i\bar{P}_{\bar{X}})/\sqrt{2\beta }%
\;\;,\;\;a_{\bar{Y}}=(\bar{X}+i\bar{P}_{\bar{X}})/\sqrt{2\beta }\;\;,
\end{equation}
for the $x$ and similarly for the $y$ direction. From these definitions we
can easily see that $[a_{x},a_{x}^{\dagger }]=1$. In terms of these
operators the non-unitary Hamiltonian takes the form 
\begin{eqnarray}
H_{non}&=&\beta \left[ (1-\delta )(a_{\bar{X}}^{\dagger }a_{\bar{X}}+a_{\bar{Y}%
}^{\dagger }a_{\bar{Y}})\right.\\\nonumber
&-&\left.\frac{\delta }{2}(a_{\bar{X}}^{\dagger 2}+a_{\bar{X}%
}^{2}+a_{\bar{Y}}^{\dagger 2}+a_{\bar{Y}}^{2})+(1-\delta )\right] \;\;,
\end{eqnarray}
Analytically evaluating the propagator $U_{non}(\tau )$ is quite tedious and
is left to Appendix 1. Inverting, $|\langle \Psi _{i}|U_{non}^{\dagger
}(\tau _{\zeta })U_{non}(\tau _{\zeta })|\Psi _{i}\rangle |^{2}=\zeta $, to
obtain $\tau _{\zeta }$ is not possible analytically and was performed
numerically using Brent's algorithm \cite{NUMERICAL_RECIPIES}.

To compute the effect of the Jump operator $C(\vec{n})$ on the state in the
Fock basis is quite involved. A number of schemes have been used previously
to increase the efficiency of the quantum trajectory method in the case
where the quantum jump is a pure momentum kick \cite{LENS,FOURIER}. These
schemes will not work in this case as the jump operator, (\ref{JUMPOP}), is
not a pure momentum kick. One could perform this jump sequentially by
applying the momentum shift in the $\bar{P}$ basis, fast Fourier
transforming (FFT) into the $\bar{X}$ basis, and then applying $\bar{X}^{2}$%
. Since we are capable of computing the jump analytically we doubt that the
FFT method would prove more efficient in this case. In the more realistic
(and more complicated) case where we do {\em not} make the harmonic
approximation, no analytical results seem to be possible and the FFT method
would be necessary. The computation of the effect of the Jump operator on
the state is given in Appendix 2.

Finally, the generation of the random direction $\vec{n}$, for the recoil of
the atom from the distribution (\ref{circular}) is explained in Appendix 3.

\subsection{Atomic Parameters}

In this section we set out the ingredients necessary for a quantum
description of the evolution of the COM. We require an atom which can be
easily cooled to significantly populate the lowest vibrational levels of the
G-L mode. The dissipation must be large enough to observe decay of the
system within the timescale of the orbital period of the atom around the
mode. If the dissipation is too weak then the dynamics  are
washed out and a secular approximation, averaging over a cycle time, would
be a good description of the evolution. This condition can be roughly
approximated by setting the recoil frequency equal to the orbital frequency
of the harmonic well, $\omega _{s}=\hbar k^{2}/2m$.

These two conditions coupled with that of the large detuning are quite
difficult to achieve. For most atoms, the ``hole'' in the beam is very dark
and the dissipation coefficient $\eta $, is very small. This makes such a
mode quite good for trapping the atom in a region of low dissipation \cite
{ERTMER_TRAP}. To achieve high dissipation $\eta $, we must make $\Gamma
/\Delta $ as large as possible. To enforce the large detuning limit we must
also have $\Delta >\Gamma $, $\Delta >\Omega _{0}$. All of these conditions
imply $\Omega _{0}\sim \Gamma \sim \Delta $. For atoms with very low Doppler
temperatures (Calcium), these conditions are hard to meet as usually $\Gamma
\ll \Omega _{0}$ (for modest beam intensities). Instead we choose to look at
Cesium, cooled to recoil. For the Cs atom, with $m=0\cdot $665$\times
10^{-25}$kg, $\lambda =657$nm, $\Gamma /2\pi =5\cdot 3$MHz, with a Gaussian-Laguerre
beam of intensity $I=4$W/m${}^{2}$, and width $w=2\times 10^{-5}$m, we get a
Rabi frequency $\Omega _{0}/2\pi \approx 5$MHz. Choosing $\Delta \sim
3\Gamma $ or $\Delta /2\pi \sim 16\cdot 4$MHz we obtain an orbital frequency
of $\omega _{s}/2\pi \sim 774$Hz. The ratio $E_{recoil}/E_{ground}\sim
2\cdot 6$, indicates a high population in the ground vibrational state of
the well. Choosing $\beta =0\cdot 25$, we have $\alpha _{x}=0\cdot 6\mu $m, 
$\alpha _{p}=0\cdot 6\times 10^{-27}$m/s, $\eta =0.0125$ and $\mu =2.310$.
For an atom cooled to recoil in a pure minimum uncertainty state (\ref
{uncertainty}) we get $\Delta \bar{P}_{r}\sim 0\cdot 58$, and $\Delta \bar{X}%
_{r}\sim 0\cdot 22$. Since the momentum shift from the bare recoil kick is, $%
\bar{P}\rightarrow \bar{P}+0\cdot 57$, and is thus on the same scale as the
quantum wavepacket we must use the full quantal master equation to describe
the evolution. For simplicity, we set the initial wavepacket to be a minimum
uncertainty state with equal variances in both $\bar{P}$ and $\bar{X}$ so
that $\Delta \bar{X}=0\cdot 35$. In the rescaled time $\tau $ the period is $%
2\pi /\beta =8\pi \approx 25$.

\subsection{Numerical Simulation}

We first simulate the dynamics of the atom initially positioned at the
centre of the trap. We plot only the mean and variance for the cartesian
position, momentum and polar angular momentum of the atom. The angular
momentum operator is given in the position representation as $\hat{L}\equiv (%
\bar{X}\bar{P}_{\bar{Y}}-\bar{P}_{\bar{X}}\bar{Y})$. In the Fock basis this
becomes $\hat{L}=i\beta (a_{\bar{X}}a_{\bar{Y}}^{\dagger }-a_{\bar{X}%
}^{\dagger }a_{\bar{Y}})/2$. The code was run on two Sun Hyper-Sparcs and
took approximately 100 CPU hours to produce $\sim $ 300 trajectories. The
results of this simulation are shown in Figure 1. From the cylindrical
symmetry we should have $\langle \bar{X}\rangle =0$ and $\langle \bar{P}_{%
\bar{X}}\rangle =0$. This is well approximated in the data as seen in Fig.
1a. In Fig. 1b the variances in both $\bar{X}$ and $\bar{Y}$ 
increase with time as the dissipation heats the system. We notice a
slight tailing off of this heating near $\tau =80$. We cannot be entirely
sure that this is not due to truncation effects as the wavepacket spreads
out to regions of large laser intensity and undergoes frequent recoil. The
plots of $\langle \bar{X}\rangle $ and $\langle \bar{Y}\rangle $ still
remain close to zero and one might suspect that this decrease in heating
rate may be a real effect. A more surprising result can be seen in Fig. 1c.
Here, we see an initial increase in $\langle \hat{L}\rangle $ and $\langle 
\hat{L}^{2}\rangle $. The variance is at all times larger than the average.
The average reaches a maximum and then decreases towards zero as $\tau $
increases beyond $\tau =50$. This is not a truncation error and is due to
the very rapid increase in variance after $\tau =50$. The diffusion of $\hat{%
L}$ becomes so fast that the average quickly drops to zero.

In the second simulation the atom is initially again in a minimum
uncertainty state but with an initial position $\bar{X}=1,\;\bar{Y}=0$, and
initial momentum $\bar{P}_{\bar{X}}=0,\;\bar{P}_{\bar{Y}}=1$. In the absence
of dissipation the atom will rotate in the $\bar{X}-\bar{Y}$ plane about the
origin with a period of $\tau _{p}=2\pi /\beta =8\pi \sim 25$. This second
simulation took longer to run since there were many more jumps per
trajectory ( to $\tau =80$ ) than in the first simulation. The reason for
this is that the initial wavepacket is in a region of higher laser intensity
and thus undergoes more frequent spontaneous emissions. Computing times grew
too lengthy on the Hyper-Sparcs. It became necessary to parallelise the code
which was then run on a multiprocessor SGI Power Challenger. The results
shown below consist of approximately 300 trajectories and correspond to 50
CPU hours on the Power Challenger. The results of the second simulation are
shown in Figure 2. We see essentially the same overall behavior as in the
first simulation. Heating is shown by the increase in the variances of $\bar{%
X}$ and $\bar{Y}$ (Fig. 2b). The variance in $\hat{L}$ again increases
faster than the mean and $\langle \hat{L}\rangle $, after gaining a maximum,
falls to zero again (Fig. 2c). In Figure 3 we have plotted the $\bar{X}-\bar{%
Y}$ probability distribution of the estimated $\rho $ at various times in
the evolution. To gauge the truncation errors we have plotted in Fig. 4a a
histogram of the number of jumps occurring within a given time interval as a
function of time. From this and the squarish probability profile in the $%
\bar{X}-\bar{Y}$ plane we see that truncation effects become significant
after $\tau =40$.

In Appendix D we solve the master equation (\ref{master2}), {\em without}
the exponential recoil kick and coupled cross terms in the dissipation. This
corresponds to semi-classical evolution (the momentum shift becoming
negligible with respect to the size of the wavepacket) of the atom 
illuminated by a coherent superposition of G-L${}_{+10}$ and G-L${}_{-10}$ modes. We see that in this case $\langle \Delta \bar{X}\rangle $
increases without bound but remains below the values obtained in the
numerical simulations. This is to be expected as there is no recoil here.
This result gives us a rough analytical check on the numerics of the
simulation. 

\section{Conclusion}
In this work we have studied the quantum dynamics of the two dimensional 
spatial COM motion of an ultra-cold Cesium atom cooled to recoil moving near 
the center of a Gaussian-Laguerre${}_{10}$ light field in the far detuned regime.
We found that the transfer of orbital angular momentum  from
the light field to the atom is mediated by spontaneous emissions. In the appropriate
limit  we find that the orbital motion of the atom
is not very sensitive to the helicity of the illuminating light field. We argue that
the helicity dependence seen in mesoscopic experiments may be due to residual
birefringence present in the material. For motion near the center of the beam
we made an harmonic approximation to the light potential. In this approximation
we solved the quantum dynamical master equation using the quantum trajectories
method. With the harmonic approximation we 
were able to solve for the non-unitary and jump evolutions analytically. This significantly increased the computational efficiency of the numerics. 
Two seperate initial
conditions were used. In the semiclassical regime, in the harmonic 
approximation and with the same initial conditions, the atom gains a net 
angular momentum. In the quantum regime, for the particular physical 
parameters
chosen, we found that the diffusion of the angular momentum was very high 
with the result that $\langle\hat{L}\rangle$ reached a maximum at a time $\tau_c$, and 
then decreased to zero. As an analytical check we solved for the quantum evolution of the atom illuminated by a superposition of 
G-L${}_{+10}$ and G-L${}_{-10}$ light in the limit of no recoil. 
The variances found were smaller than those seen in the fully quantum computation as expected. 
From similiar studies in other systems one might expect that $\tau_c$  will increase
 with an 
increase in the initial state energy or a decrease in the dissipation rate. However, to explore the parameter space of this system more fully and include the complete exponential character of the light field potential
would require
a very significant increase in computational resources. 
With the inclusion of
the exponential in the potential and dissipation
terms of the master equation we would expect to see shear
in the evolution of the wavepacket. Partial revivals and cat states might be
expected as these have been seen in other nonlinear potentials.
We note that unless some new analytical techniques are found further investigations into this system will be very computationally expensive.

Finally,
experimental probing of $\langle \hat{L}\rangle $, for ultra-cold atoms in
the center of a Gaussian-Laguerre trap could be attempted using the azimuthal Doppler
shift provided by the rotation of the atoms about the axis of the beam \cite
{AZIMUTHAL}. One might double the shift by retro-reflecting the laser probe
beam back through the rotating atomic cloud while maintaining the reflected
beam's impact parameter to the  axis to be the same as the incident
beam's impact parameter. In order to achieve a good Doppler signal it may be
necessary to work in a regime where the orbital frequency of the atomic
cloud near the center of the  G-L mode is higher than that used in this work.

\section*{Acknolwedgements}

We thank S. Dyrting, G. Milburn, H. Wiseman, J. Steinbach and D. Segal for helpful
comments on this work. This work was made possible through a Human Capital
and Mobility Fellowship from the European Union.

\appendix

\section{Matrix elements for $U_{non}$}

In this appendix we calculate the matrix elements of the non-unitary
propagator $U_{non}$ in the number state basis. Although the following
calculations are somewhat involved the resulting analytic expression proved
more efficient than a direct numerical integration of the equations. We
begin with the general Fock state basis matrix element 
\begin{eqnarray}
&&\langle m_{x},m_{y}| e^{-iH_{non}\tau /\beta }|n_{x},n_{y}\rangle = 
\nonumber \\
&&e^{-i\tau (1-\delta )}\langle m_{x}|\exp \left\{ -i\tau \left[ (1-\delta
)a_{x}^{\dagger }a_{x}-\frac{\delta }{2}(a_{x}^{\dagger 2}+a_{x}^{2})\right]
\right\} |n_{x}\rangle  \nonumber \\
&&\times \langle m_{y}|\exp \left\{ -i\tau \left[ (1-\delta )a_{y}^{\dagger
}a_{y}-\frac{\delta }{2}(a_{y}^{\dagger 2}+a_{y}^{2})\right] \right\}
|n_{y}\rangle
\end{eqnarray}
where $\delta =i\beta \eta $.

To evaluate this we use Baker-Campbell-Hausdorff (BCH) disentangling. The
particular methods is due to Wilkox \cite{WILKOX}. For simplicity we take
first the $x$ matrix elements and drop the $x$ subscript. We then write $%
U_{non}=\exp\, ( {\cal H})$ where ${\cal H}\equiv xK_0+y(K_++K_-)$, and
where $K_0,K_\pm$ are the generators of the Lie group SU(1,1) which satisfy 
\begin{equation}
K_0\equiv\frac{1}{2}(a^\dagger a+1/2)\;\;,\;\;K_+\equiv\frac{1}{2}a^{\dagger
2}\;\;, \;\;K_-\equiv \frac{1}{2}a^2\;\;,
\end{equation}
\begin{equation}
[K_+,K_-]=-K_0\;\;,\;[K_0,K_\pm]=\mp 2K_0\;\;,
\end{equation}
\begin{equation}
x=-2i\tau(1-\delta)\;\;,\; y=+i\tau\delta\;\;.
\end{equation}
To evaluate the matrix element between number states we must first re-write $%
U_{non}$ in normal order. To do this we use the parameter differential
methods of Wilkox \cite{WILKOX}. Set ${\cal U}(\epsilon )=\exp \,({\cal H}%
\epsilon)$, this gives 
\begin{equation}
\frac{d{\cal U}}{d\epsilon}={\cal HU}\;\;,  \label{A1.1}
\end{equation}
and put ${\cal H}=a_+K_++a_0K_0+a_-K_-$, ($x=a_0$ and $y=a_-=a_+$). Adopting
the ansatz 
\begin{equation}
{\cal U}\equiv
e^{g_+(\epsilon)K_+}\,e^{g_0(\epsilon)K_0}\,e^{g_-(\epsilon)K_-}\;\;,
\label{A1.6}
\end{equation}
we perform the differentiation on the left hand side of (\ref{A1.1}) and use
BCH disentangling to obtain ordinary differential equations relating the $%
g_i(\epsilon)$ to the $a_i$. The solution of these ODEs for $\epsilon=1$ and
initial $g_i(\epsilon=0)$ gives the required normal ordering.

Differentiating ${\cal U}(\epsilon)$ gives 
\begin{eqnarray}
&&\frac{d{\cal U}}{d\epsilon} ={\cal HU}  \nonumber \\
& &= \left(\dot{g}_+K_++\dot{g}_0\,e^{g_+ {\rm ad}K_+}K_0+\dot{g}_-\,e^{g_+%
{\rm ad}K_+}\,e^{g_0{\rm ad}K_0}K_-\right){\cal U}  \nonumber \\
& & =\left[a_+K_++a_-K_-+a_0K_0\right]{\cal U}  \label{A1.2}
\end{eqnarray}

We now use the commutation properties of SU(1,1) and the BCH formula 
\begin{eqnarray}
e^{\xi A}\,B\,e^{-\xi A}=B+\xi [A,B]&+&\frac{\xi ^{2}}{2!}[A,[A,B]]\\\nonumber
&+&\frac{\xi
^{3}}{3!}[A,[A,[A,B]]]+\cdots
\end{eqnarray}
($A$ and $B$ are operators while $\xi $ is a scalar) to construct the
adjoint action table (Table A1) 

Using this information we can evaluate the adjoint actions in equation (\ref
{A1.2}) giving 
\begin{eqnarray}
&&\left\{ K_{+}\left( \dot{g}_{+}-2g_{+}\dot{g}_{0}+e^{-2g_{0}}g_{+}^{2}\dot{%
g}_{-}\right) +K_{0}\left( \dot{g}_{0}
-g_{+}e^{-2g_{0}}\dot{g}_{-}\right)\right.\nonumber\\\nonumber
&+&\left.K_{-}\left( e^{-2g_{0}}\dot{g}_{-}\right) \right\}\nonumber   \\\nonumber
&=&a_{+}K_{+}+a_{-}K_{-}+a_{0}K_{0}\;\;.
\end{eqnarray}
Equating the coefficients of $K_{\pm }\;,\;\;K_{0}$, and re-arranging gives 
\begin{eqnarray}
\dot{g}_{-} &=&a_{-}e^{2g_{0}}\;\;, \\
\dot{g}_{0} &=&a_{0}+g_{+}a_{-}\;\;, \\
\dot{g}_{+} &=&g_{+}(g_{+}a_{-}+2a_{0})+a_{+}\;\;.
\end{eqnarray}
Solving these with the initial condition $g_{i}(\epsilon =0)=0$ and setting $%
\epsilon =1$ in the final result we get 
\begin{equation}
g_{+}=\frac{a_{+}\tan \gamma }{\gamma -a_{0}\tan \gamma }\;\;,\;\;g_{-}=%
\frac{a_{-}\tan \gamma }{\gamma -a_{0}\tan \gamma }\;\;,\end{equation}
\begin{equation}g_{0}=-\ln
\left[ \cos \gamma -\frac{a_{0}}{\gamma }\sin \gamma \right] \;\;,
\end{equation}
where $\gamma ^{2}=a_{+}a_{-}-a_{0}^{2}=(4+3\delta ^{2}-8\delta )\tau ^{2}$.
To obtain the matrix element of the normally ordered operator, (ie. $\langle
m|{\cal U}|n\rangle \rangle $) we can proceed in a number of ways, however
the most straightforward is to insert resolutions of unity in the coherent
state representation, 
\begin{equation}
\langle m|{\cal U}|n\rangle =\int \frac{d^{2}\alpha }{\pi }\int \frac{%
d^{2}\beta }{\pi }\,\langle m|\alpha \rangle \langle \alpha |{\cal U}|\beta
\rangle \langle \beta |n\rangle \;\;.
\end{equation}
From (\ref{A1.6}) we get 
\begin{eqnarray}
\langle m|{\cal U}|n\rangle& =&\langle \alpha
|e^{g_{+}K_{+}}\,e^{g_{0}K_{0}}\,e^{g_{-}K_{-}}\,|\beta \rangle \;\; 
\nonumber \\
&=&\exp \left\{ \frac{1}{4}g_{0}+\frac{1}{2}g_{+}\alpha ^{*2}+\frac{1}{2}%
g_{-}\beta ^{2}+e^{g_{0}/2}\alpha ^{*}\beta\right.\nonumber\\
& &\;\;\;\qquad\left. -\frac{1}{2}(|\alpha
|^{2}+|\beta |^{2})\right\}  \nonumber \\
&\equiv& {\cal {\bf D}}\;\;,
\end{eqnarray}
where we have used $\exp \,(g_{0}a^{\dagger }a/2)=\sum_{0}^{\infty }\,(\exp
\,(g_{0}/2)-1)^{l}a^{\dagger l}a^{l}l!^{-1}$. The normally ordered element
can now be expressed as 
\begin{eqnarray}
\langle m|{\cal U}|n\rangle &=&\frac{1}{\pi ^{2}}\int \frac{d^{2}\alpha
d^{2}\beta }{\sqrt{n!m!}}\,\alpha ^{n}\beta ^{*m}\,{\cal {\bf D}}  \nonumber
\\
&\equiv &\left. \frac{d^{n}}{d\gamma _{1}^{n}}\frac{d^{m}}{d\gamma _{2}^{m}}%
\,{\bf \Pi }\right| _{\gamma _{1}=0,\,\gamma _{2}=0}\;\;,  \label{A1.diff}
\end{eqnarray}
where \\
\vspace{1cm}\\
\noindent\rule{0.5\textwidth}{0.4pt}\rule{0.4pt}{\baselineskip}
\begin{eqnarray}
{\bf \Pi } &=&\frac{1}{\pi ^{2}}\int \frac{d^{2}\alpha d^{2}\beta }{\sqrt{%
n!m!}}\exp \left\{ -\frac{1}{2}(|\alpha |^{2}+|\beta
|^{2})+e^{g_{0}/2}\alpha ^{*}\beta\right.\nonumber\\
&&\;\;\;\;\;\;\qquad\left. +\frac{1}{2}g_{+}\alpha ^{*2}+\frac{1}{2}%
g_{-}\beta ^{2}+\gamma _{1}\alpha +\gamma _{2}\beta ^{*}+\frac{1}{4}%
g_{0}\right\}  \nonumber \\
&=&\frac{1}{\sqrt{n!m!}}\exp \,\left( \frac{\gamma _{2}^{2}}{2}g_{-}+\frac{%
\gamma _{1}^{2}}{2}g_{+}+\gamma _{1}\gamma _{2}\,e^{g_{0}/2}\right)
\,e^{g_{0}/4}\;\;.
\end{eqnarray}
To perform the differentiations in (\ref{A1.diff}) we make use of the
identities 
\begin{equation}
H_{n}{x}=(-1)^{n}e^{x^{2}}\frac{d^{n}}{dx^{n}}\,e^{-x^{2}}\;\;,\;%
\;n!e^{-x}x^{\alpha }L_{n}^{\alpha }(x)=\frac{d^{n}}{dx^{n}}%
\,(e^{-x}x^{n+\alpha })\;\;,
\end{equation}
where $H_{n},\;L_{n}^{\alpha }$ are the Hermite and associated Laguerre
polynomials. Writing $\Pi =\exp (g_{0}/4)\exp (ax^{2}+by^{2}+cxy)/\sqrt{n!m!}
$ where $a=g_{+}/2\;,b=g_{-}/2$ and $c=\exp \,(g_{0}/2)$ one can, after
expanding products, differentiating and setting $\gamma _{1}=\gamma _{2}=0$,
obtain 
\end{multicols}
\begin{eqnarray}
\langle m|{\cal U}|n\rangle &=&  \nonumber \\
&&\frac{e^{g_{0}/4}}{\sqrt{n!m!}}\sum_{j=-min(0,\Delta )}^{[n/2]}\,e^{-i\pi
(j+k)}a^{j}b^{k}c^{n-2j}(m-2k)!H_{2j}(0)H_{2k}(0)\left( 
\begin{array}{c}
n \\ 
2j
\end{array}
\right) \left( 
\begin{array}{c}
m \\ 
2k
\end{array}
\right) \;\;,  \label{A1.Horrible}
\end{eqnarray}
where $k=j-\Delta $, $\Delta =(n-m)/2$, and $[n/2]$ represents the integer
part of $n/2$. To construct the matrix elements for the combined $x-y$
system we have $|\Psi \rangle =\sum \,A(n_{x},n_{y})|n_{x},n_{y}\rangle $.
Under the non-unitary evolution this becomes $U_{non}(\tau )|\Psi \rangle
=|\Psi ^{\prime }\rangle =\sum \,B(m_{x},m_{y})|m_{x},m_{y}\rangle $ where 
\begin{eqnarray}
B(m_{x},m_{y}) &=&\langle m_{x},m_{y}|U_{non}(\tau
)|\sum_{n_{x},n_{y}}\,A(n_{x},n_{y})|n_{x},n_{y}\rangle \;\;,  \nonumber \\
&=&\sum_{n_{x},n_{y}}\,A(n_{x},n_{y})\langle m_{x}|{\cal U}(\tau
)|n_{x}\rangle \langle m_{y}|{\cal U}(\tau )|n_{y}\rangle \;\;.
\end{eqnarray}
Denoting the matrix element of ${\cal U}$ by ${\cal U}_{ab}$, we have the
simple relation $B_{m_{x},m_{y}}=({\cal U}\cdot A\cdot {\cal U}^{{\rm T}%
})_{m_{x},m_{y}}$. This completes the analytic description of the
non-unitary propagator. We note that the main numerical overhead occurs in
the computation of the sum in (\ref{A1.Horrible}) for a given value of $\tau 
$. Lookup tables for the factorial and Hermite functions were used to
increase efficiency.

\section{Matrix elements for the Jump operation}

In this appendix we calculate the matrix elements for the Jump operation. To
jump we essentially apply the operator 
\begin{equation}
\hat{C}=(\hat{x}+i\hat{y})e^{i\mu (\epsilon _{x}\hat{x}+\epsilon _{y}\hat{y}%
)}\;\;,
\end{equation}
to the state $|\Psi \rangle $ where $\epsilon _{x,y}$ are the direction
cosines for the recoil kick and $\mu $ is related to the momentum
transferred. We are thus reduced to finding the matrix element 
\begin{equation}
\langle m_{x},m_{y}|\,(\hat{x}+i\hat{y})e^{i\mu (\epsilon _{x}\hat{x}%
+\epsilon _{y}\hat{y})}\,|n_{x},n_{y}\rangle \;\;.
\end{equation}
We begin by evaluating $\langle m|\exp (ib\hat{x})|n\rangle $ and then
differentiating with respect to $b$. From section (II) we have $\bar{x}=%
\sqrt{\beta /2}(a^{\dagger }+a)$ and putting $\kappa =b\sqrt{\beta /2}$ and
using the definition of $|n\rangle $ with BCH disentangling we obtain 
\begin{equation}
\langle m|e^{ib\bar{x}}|n\rangle =\frac{1}{\sqrt{n!m!}}\langle
0|a^{m}e^{i\kappa a^{\dagger }}e^{i\kappa a}a^{\dagger n}|0\rangle
e^{-\kappa ^{2}/2}\;\;.
\end{equation}
We can pull the annihilation operators to the right using 
\begin{equation}
a^{m}e^{i\kappa a^{\dagger }}=e^{i\kappa a^{\dagger }}e^{-i\kappa a^{\dagger
}}\,a^{m}e^{i\kappa a^{\dagger }}=e^{i\kappa a^{\dagger }}(a+i\kappa
)^{m}\;\;.
\end{equation}
Similarly, $\exp (i\kappa a)a^{\dagger n}=(a^{\dagger }+i\kappa )^{n}\exp
(i\kappa a)$. We thus must evaluate 
\begin{equation}
\langle m|e^{ib\bar{x}}|n\rangle =\frac{e^{-\kappa ^{2}/2}}{\sqrt{n!m!}}%
\langle 0|e^{i\kappa a^{\dagger }}(a+i\kappa )^{m}(a^{\dagger }+i\kappa
)^{n}e^{i\kappa a}|0\rangle \;\;.  \label{A2.1}
\end{equation}
Binomially expanding the terms in (\ref{A2.1}) we finally obtain 
\begin{eqnarray}
\langle m|e^{ib\bar{x}}|n\rangle &=&\sqrt{\frac{n_{>}!}{n_{<}!}}e^{-\kappa
^{2}/2}\sum_{j=0}^{n_{<}}\,\left( 
\begin{array}{c}
n_{<} \\ 
j
\end{array}
\right) \frac{(i\kappa )^{n_{>}+n_{<}-2j}}{(n_{>}-j)!}\;\;,  \nonumber \\
&=&\sqrt{\frac{n_{>}!}{n_{<}!}}e^{-\kappa ^{2}/2}\frac{(i\kappa )^{n+m}}{%
n_{>}!}{\bf F}([-n_{>},-n_{<}],0,-\frac{1}{\kappa ^{2}})\;\;, \\
&\equiv &{\cal G}(m,n,b)\;\;,
\end{eqnarray}
where $n_{>}={\rm max}(n,m)$,$n_{<}={\rm min}(n,m)$ and ${\cal F}$ is the
generalised hypergeometric function \cite{HYPERGEO}. Now to obtain the
matrix element $\langle m|e^{ib\bar{x}}|n\rangle $ we differentiate with
respect to $\kappa $ to get 
\begin{eqnarray}
\langle m|e^{ib\bar{x}}|n\rangle &=&-i\sqrt{\frac{\beta }{2}}\frac{d}{%
d\kappa }\langle m|e^{i\kappa (a+a^{\dagger })}|n\rangle  \nonumber \\
&=&-i\sqrt{\frac{\beta }{2}}\sqrt{\frac{n_{>}!}{n_{<}!}}\frac{e^{-\kappa
^{2}/2}(i\kappa )^{n+m}}{\kappa ^{3}n_{>}!}\left\{ -\kappa ^{4}{\bf F}%
([-n_{>},-n_{<}],0,-\frac{1}{\kappa ^{2}})\right.  \nonumber \\
&+&\left. \kappa ^{1}(n+m){\bf F}([-n_{>},-n_{<}],0,-\frac{1}{\kappa ^{2}}%
)+2n_{>}n_{<}{\bf F}([-n_{>}+1,-n_{<}+1],0,-\frac{1}{\kappa ^{2}})\right\}
\;\;, \\
&\equiv &{\cal F}(m,n,b)\;\;.
\end{eqnarray}
The complete two dimensional matrix element may now be expressed as 
\begin{eqnarray}
\langle m_{x},m_{y}| &(\bar{x}+i\bar{y})&e^{i\mu (\epsilon _{x}\bar{x}%
+\epsilon _{y}\bar{y})}|n_{x},n_{y}\rangle  \nonumber \\
&=&{\cal F}(m_{x},n_{x},\mu \epsilon _{x}){\cal G}(m_{y},n_{y},\mu \epsilon
_{y})+i{\cal G}(m_{x},n_{x},\mu \epsilon _{x}){\cal F}(m_{y},n_{y},\mu
\epsilon _{y})  \nonumber \\
&\equiv &{\cal T}(m_{x},m_{y};n_{x},n_{y})\;\;.
\end{eqnarray}
Denoting the pre-jump state by $|\Psi \rangle =\sum
\,A(n_{x},n_{y})|n_{x},n_{y}\rangle $, then $|\langle \Psi |\Psi \rangle
|^{2}<1$ after then jump. Letting the state after the jump be $|\Psi \rangle
_{J}=\hat{C}|\Psi \rangle =\sum \,B(m_{x},m_{y})|m_{x},m_{y}\rangle $ then
the coefficients $B(m_{x},m_{y})$ are given by 
\begin{equation}
B(m_{x},m_{y})=\sum_{n_{x},n_{y}}\,A(n_{x},n_{y}){\cal T}%
(m_{x},m_{y};n_{x},n_{y})\;\;.
\end{equation}
The main numerical overhead occurs in the calculation of ${\cal G}$ and $%
{\cal F}$. Since the jump directions are chosen randomly, the values for
these functions must be computed each time. However, lookup tables for the
generalised hypergeometric function are used and only two evaluations of $%
{\bf F}$ are needed for each jump computation. After a jump the state is
renormalised.

\hspace*{\fill}\rule[0.4pt]{0.4pt}{\baselineskip}%
\rule[\baselineskip]{0.5\textwidth}{0.4pt}  
\begin{multicols}{2}
\section{Generation of Random Emission Vectors}

To generate the random direction vectors, $\vec{n}(\phi ,\theta )$, sampled
from the probability distribution (\ref{circular}) we first note that the
probability is independent of the angle $\phi $. Thus we choose $\phi $ to
be a uniform random variable $\phi \in [0,2\pi ]$. To generate the $\theta $%
, a random variable sampled from the probability distribution $1+\cos
^{2}\theta $, we use the method of cumulative inversion. We set the
cumulative probability in $\theta $ equal to a uniform random variable and
invert to obtain $\theta $, 
\begin{equation}
\epsilon =\int_{0}^{\pi }\,\sin \theta \,d\theta \,\frac{3}{16\pi }(1+\cos
^{2}\theta )\;\;,
\end{equation}
gives 
\begin{equation}
\theta =\cos ^{-1}\left( u^{1/3}-u^{-1/3}\right) \;\;,
\end{equation}
where $u=2-4\epsilon +\sqrt{5-16\epsilon +16\epsilon ^{2}}$.

\section{Caldeira-Leggett Model}

The dynamics of an atom moving in the semiclassical regime ($\mu =0$), near
the beam axis for a beam in a superposition of the
G-L${}_{+10}$ and G-L${}_{-10}$ is that of the Caldeira-Leggett model \cite
{HALLIWELL_DOWKER}. This mode does not possess any orbital angular
momentum and thus there are no cross terms in the dissipation in the quantum
master equation. The equation separates into $x$ and $y$ components. In the
position representation the $x$-component equation is 
\begin{equation}
i\beta \frac{d}{d\tau }\rho =\left[ -\frac{\beta ^{2}}{2}\left( \partial _{%
\bar{x}}^{2}-\partial _{\bar{x}^{\prime }}^{2}\right) +\frac{1}{2}(\bar{x}%
^{2}-\bar{x}^{\prime 2})-i\bar{\eta}\beta (\bar{x}-\bar{x}^{\prime
2})\right] \rho \;\;.
\end{equation}
In \cite{HALLIWELL_DOWKER}, the master equation, in the limits 
\begin{equation}
\gamma \rightarrow 0\;\;,\;\;T\rightarrow \infty \;\;,\;\;k_{B}\gamma
T\rightarrow \eta \;\;,  \label{limits}
\end{equation}
is 
\begin{equation}
i\frac{d}{dt}\rho =\left[ -\frac{1}{2}\left( \partial _{x}^{2}-\partial
_{x^{\prime }}^{2}\right) +\frac{\omega _{R}^{2}}{2}(x^{2}-x^{\prime
2})-i\eta (x-x^{\prime })^{2}\right] \rho \;\;.
\end{equation}
To convert between these two equations we set $\tau =\beta t$, $\bar{x}%
=\beta x$ and $\eta =\bar{\eta}\beta ^{3}$. This gives $\omega
_{R}^{2}=\beta ^{2}$ and variance, $\langle \bar{x}^{2}\rangle =\beta
^{2}\langle x^{2}\rangle $. In the limit (\ref{limits}), the coefficients
(in \cite{HALLIWELL_DOWKER}, equations 4.41, 4.42, 4.43) become 
\begin{eqnarray}
A &=&\frac{\eta }{2\omega _{R}\sin ^{2}\nu }\left[ \nu -\sin 2\nu \right]
\;\;, \\
B &=&\frac{\eta }{\omega _{R}\sin ^{2}\nu }\left[ \sin \nu -\nu \cos \nu
\right] \;\;, \\
C &=&\frac{\eta }{2\omega _{R}\sin ^{2}\nu }\left[ \nu -\frac{1}{2}\sin 2\nu
\right] \;\;,
\end{eqnarray}
where $\tilde{K}=\hat{K}=\frac{\omega _{R}}{2\cot \nu }$, $L=N=\frac{\omega
_{R}}{2\sin \nu }$ and where $\nu =\omega _{R}t$. We begin with the density
matrix (6.14) in \cite{HALLIWELL_DOWKER} we can use their solution for the
evolved $\rho (t)$ in (6.15), 
\begin{equation}
\rho (t)=\tilde{N}^{2}\exp \,\left( -{\cal A}\zeta _{1}^{2}-{\cal B}%
X_{1}^{2}+i\zeta _{1}X_{1}{\cal C}\right) \;\;,
\end{equation}
where $X_{1}=X+X^{\prime }$, $\zeta _{1}=X-X^{\prime }$ , are the sum and
relative coordinates for the evolved density matrix. Now with the initial
condition $\langle P\rangle =0$ , we have $\langle x^{2}\rangle =\frac{1}{8%
{\cal B}}$ for any $t$. So we just proceed to carefully evaluate the
coefficient ${\cal B}$. Beginning with an initial state in the barred
variables 
\begin{equation}
\rho _{A}=\frac{1}{\sqrt{2}\pi \sigma ^{2}}\,e^{-\frac{(X_{0}^{2}+\zeta
_{0}^{2})}{8\sigma ^{2}}}\;\;,
\end{equation}
where $X_{0}=X+X^{\prime }$ and $\zeta _{0}=X-X^{\prime }$, the coordinates
for the initial state. Transforming to the barred variables we have $\bar{%
\sigma}=\sigma \beta $. Collecting all of the above we get 
\begin{eqnarray}
\langle \bar{x}^{2}\rangle &=&\langle x^{2}\rangle \beta ^{2}  \nonumber \\
&=&\frac{\beta ^{2}}{8{\cal B}}  \nonumber \\
&=&\bar{\sigma}^{2}\cos ^{2}\nu +\bar{\eta}\beta ^{2}[\nu -\frac{1}{2}\sin
2\nu ]+\frac{\beta ^{2}}{4\bar{\sigma}^{2}}\sin ^{2}\nu \;\;,
\end{eqnarray}
where $\nu =\omega _{R}t=\beta t=\tau $. Plotting this in Figure 6 for the
initial conditions $\bar{\eta}=0.0125$, $\beta =.25$ and $\bar{\sigma}=0.35$
we see that the variance generally increases with time $\tau $. It remains
smaller than the variance calculated numerically for the G-L${}_{10}$ mode
with recoil.

\end{multicols}
\begin{table}[tbp]
\caption{Table of adjoint action of SU(1,1): $e^{x\Gamma _{g}}\,\Gamma
_{k}\,e^{-z\Gamma _{j}}$}\squeezetable
\begin{tabular}{|c||c|c|c|}
$\Gamma_j\backslash \Gamma_k$ & $K_0$ & $K_+$ & $K_-$ \\ \hline\hline
$K_0$ & $K_0$ & $e^{2x}K_+$ & $e^{-2x}K_-$ \\ \hline
$K_+$ & $K_0-2xK_+$ & $K_+$ & $K_--xK_0+x^2K_+$ \\ \hline
$K_-$ & $K_0+2xK_-$ & $K_++xK_0+x^2K_-$ & $K_-$%
\end{tabular}
\end{table}
%
%
%
%
\newpage
\setlength{\oddsidemargin}{-.5in}
\setlength{\textwidth}{8in}
\begin{multicols}{2}
\begin{minipage}{3.25in}\flushleft
\begin{figure}
\begin{center}
\setlength{\unitlength}{1.cm}
\epsfig{file=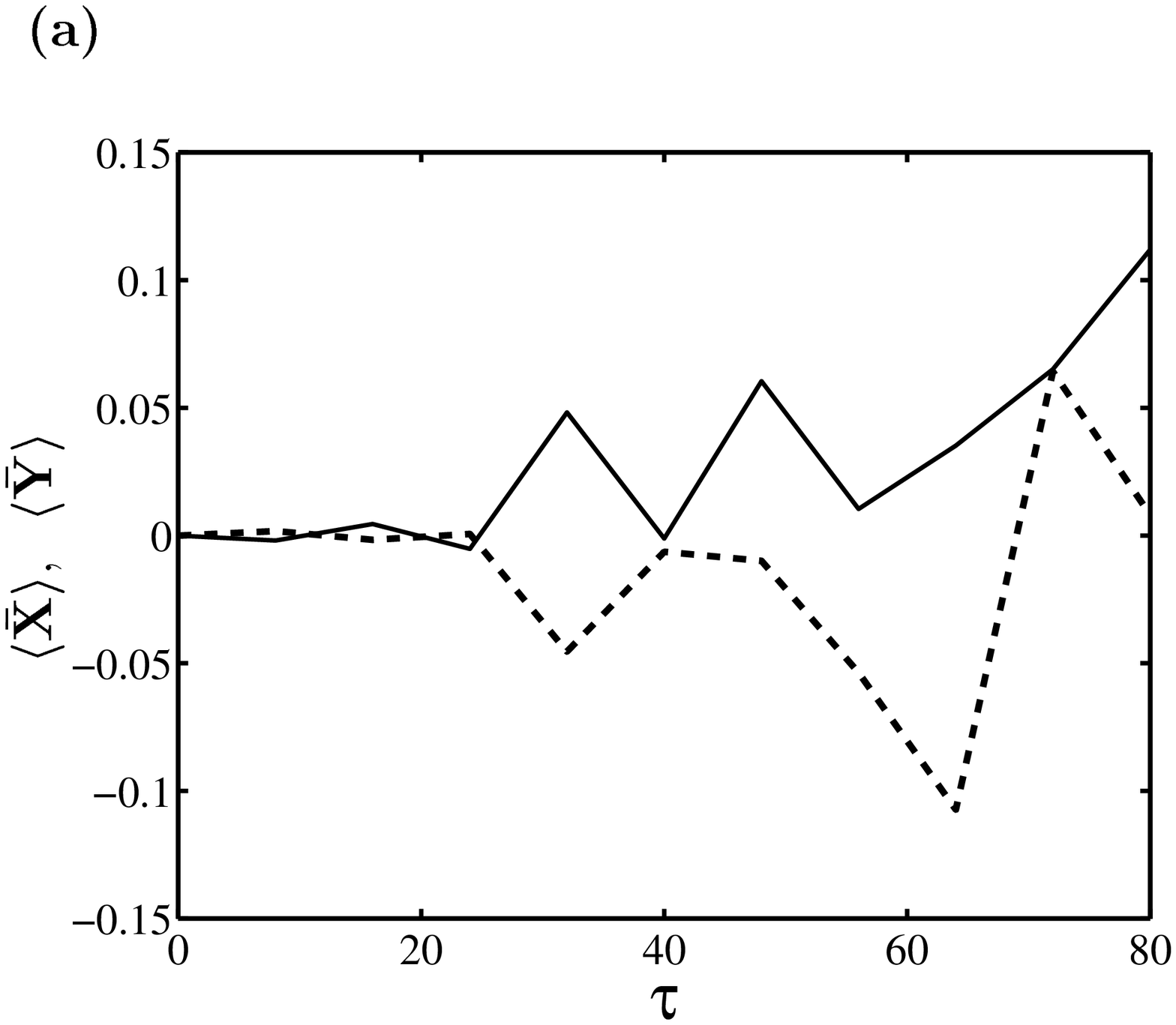,width=7cm}\\ 
\epsfig{file=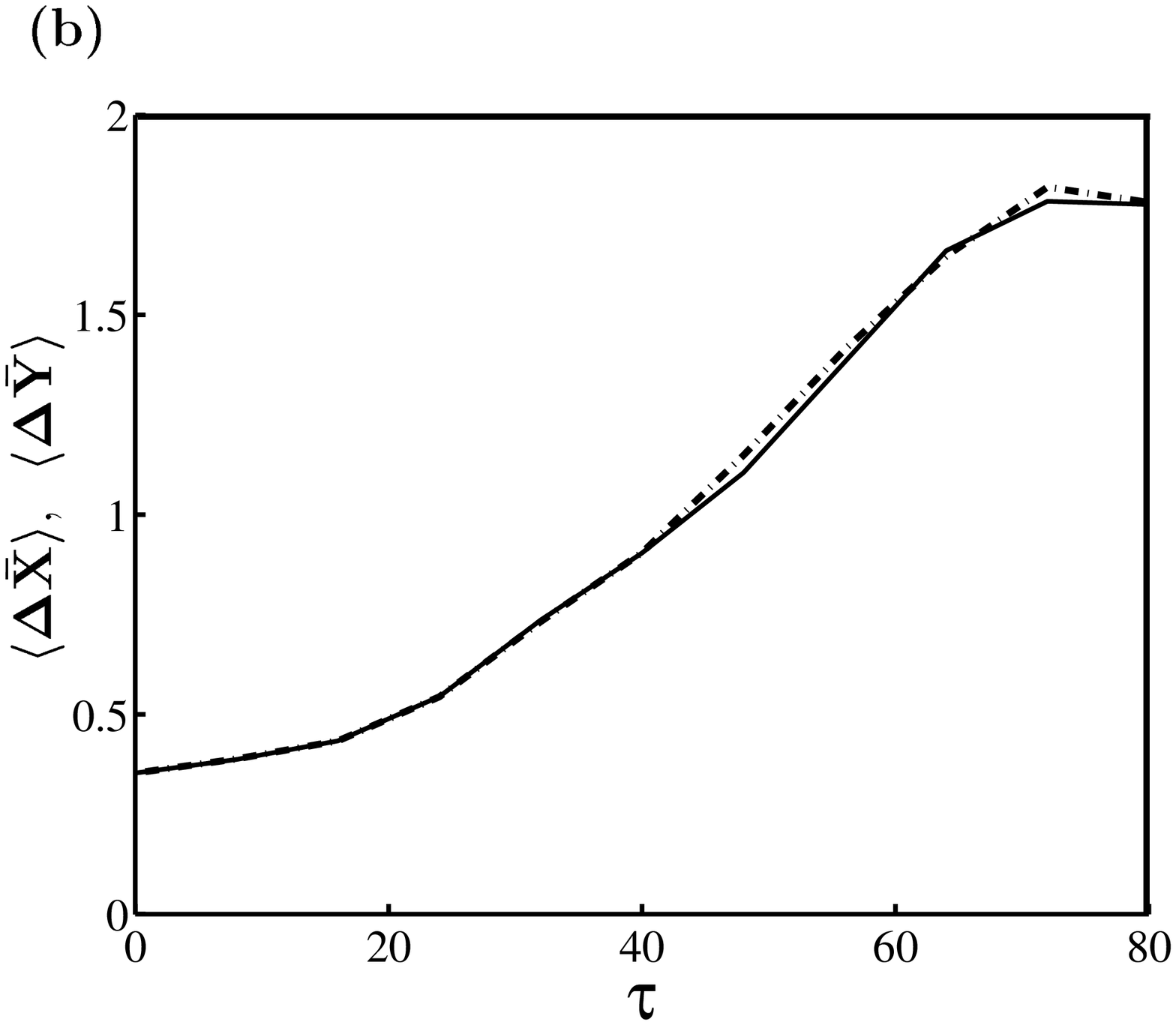,width=7cm}\\ 
\epsfig{file=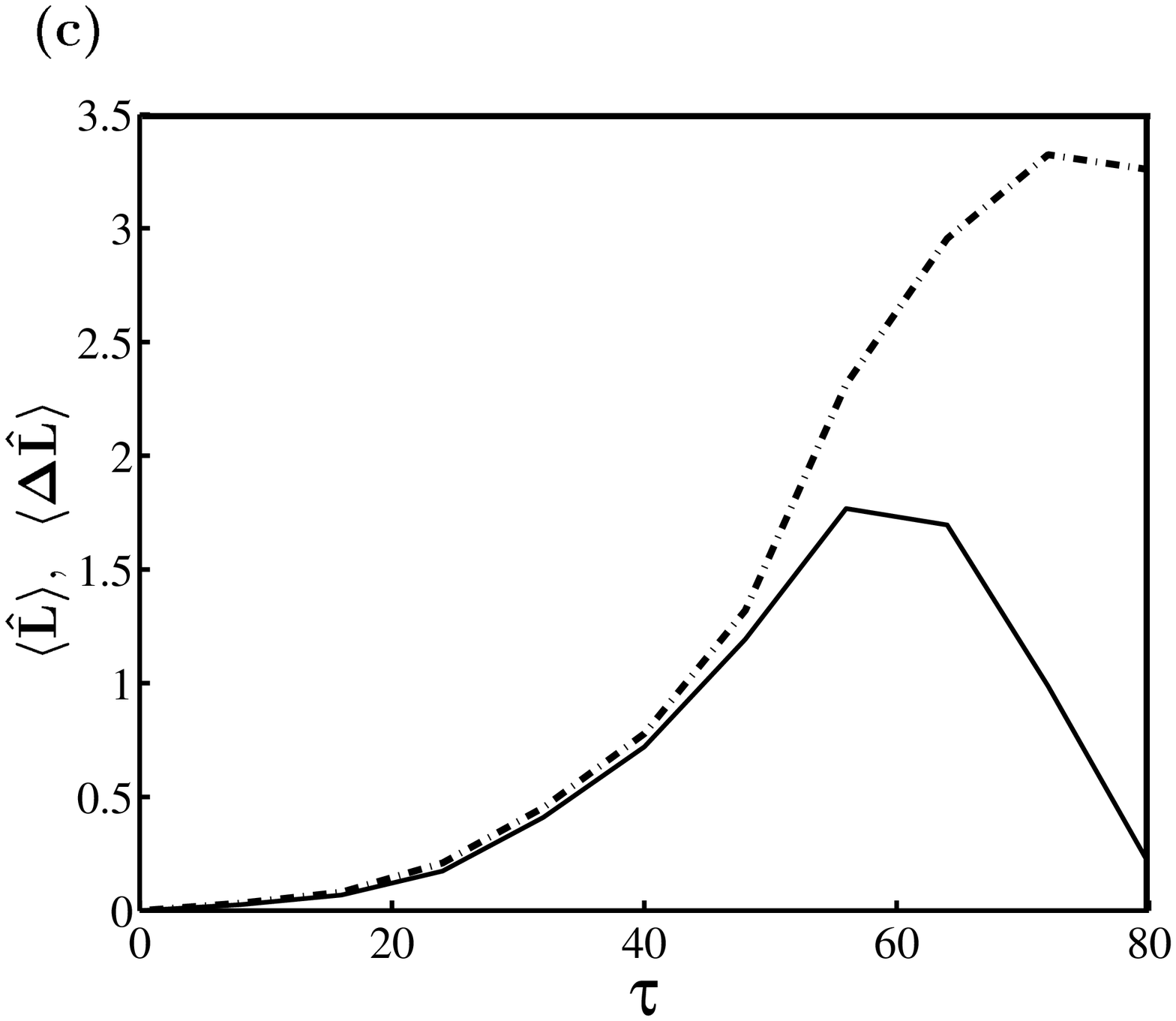,width=7cm}
\end{center}
\caption{Plots of (a) $\langle \bar{X}\rangle $, $\langle \bar{Y}\rangle $,
(b) $\langle \Delta \bar{X}\rangle $, $\langle \Delta \bar{Y}\rangle $, ($%
\bar{X}$ solid, $\bar{Y}$ dashed), and (c) $\langle \hat{L}\rangle $, $%
\langle \Delta \hat{L}\rangle $, (solid, dashed), for Simulation 1 where the
initial state is a minimum uncertainty state at the origin.}
\end{figure}
\end{minipage}
\begin{minipage}{3.25in}
\begin{figure}
\begin{center}
\setlength{\unitlength}{1.cm}
\epsfig{file=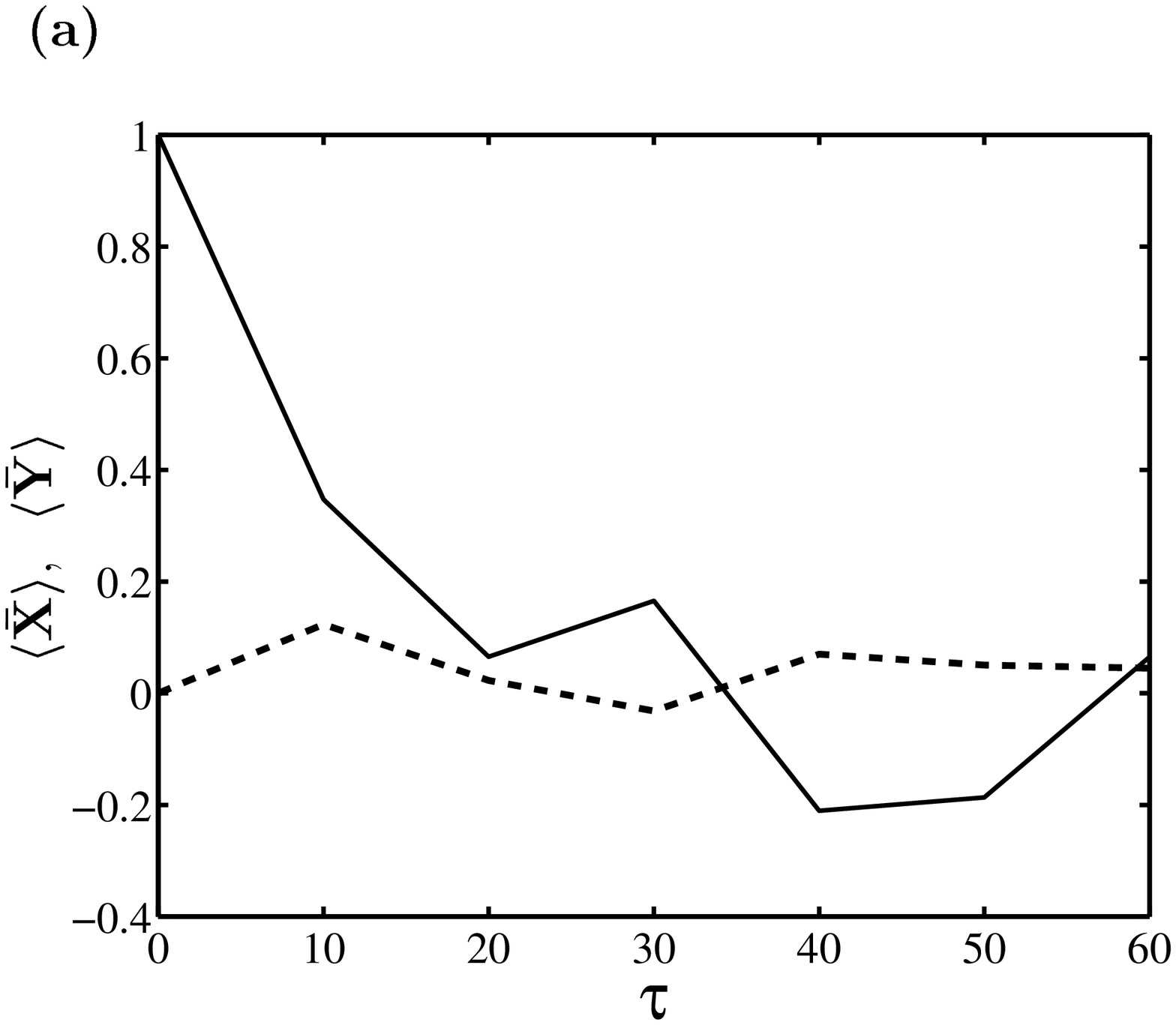,width=7cm}\\ 
\epsfig{file=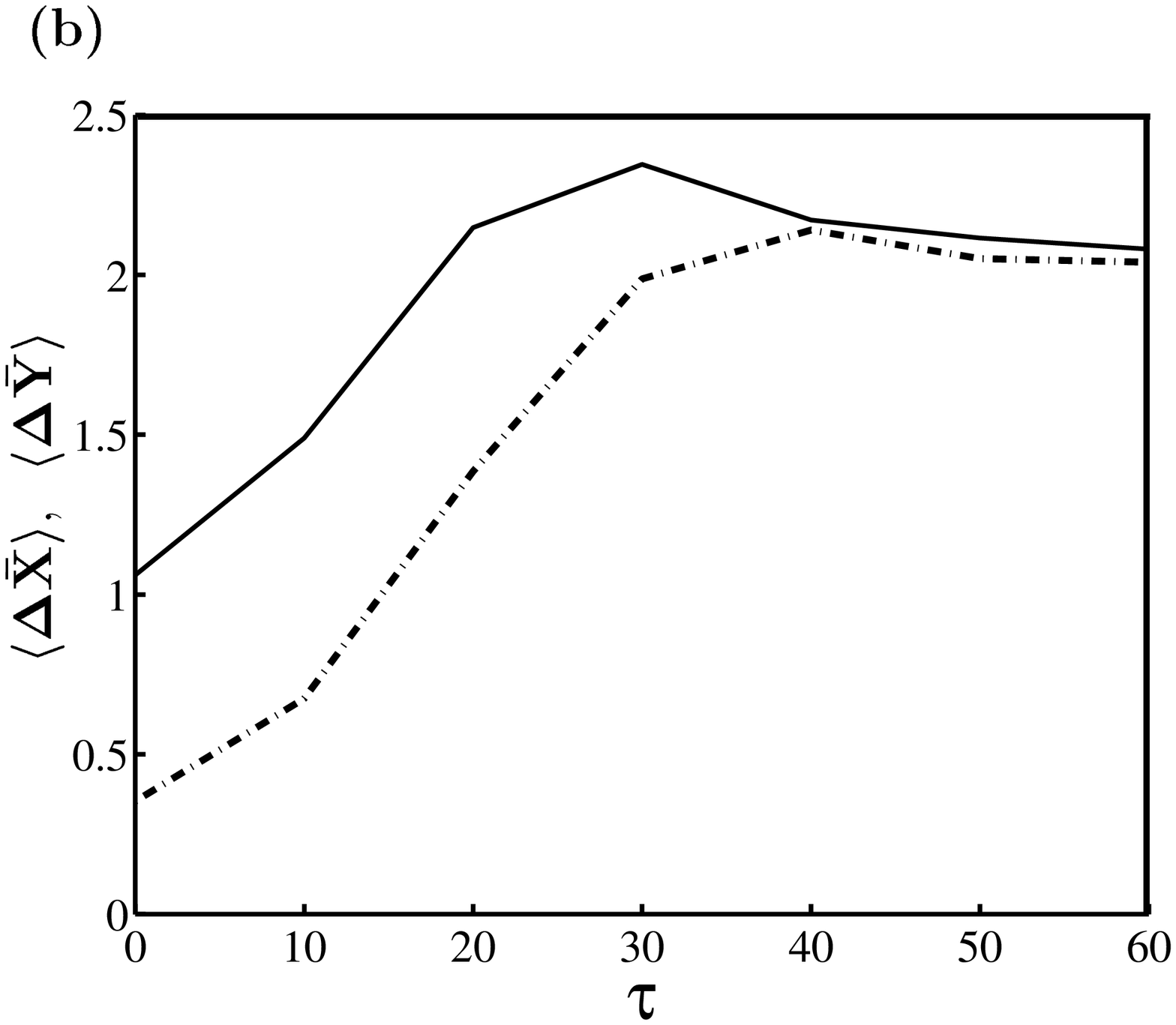,width=7cm}\\
\epsfig{file=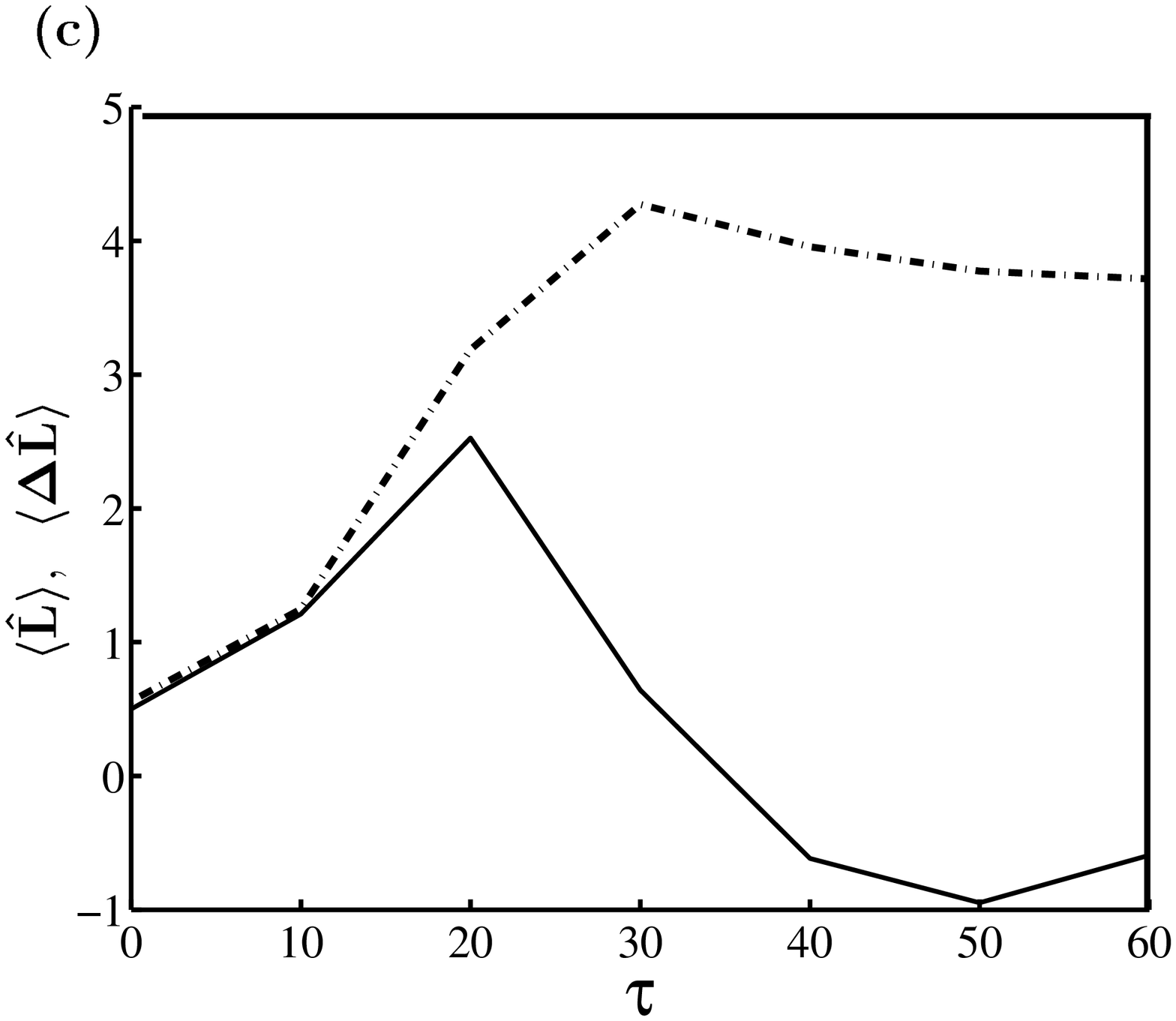,width=7cm}
\end{center}
\caption{Plots of (a) $\langle \bar{X}\rangle $, $\langle \bar{Y}\rangle $,
(b) $\langle \Delta \bar{X}\rangle $, $\langle \Delta \bar{Y}\rangle $, ($%
\bar{X}$ solid, $\bar{Y}$ dashed), and (c) $\langle \hat{L}\rangle $, $%
\langle \Delta \hat{L}\rangle $, (solid, dashed), for Simulation 2 where the
initial state is a minimum uncertainty state offset from the origin.}
\end{figure}
\end{minipage}
\end{multicols}
\newpage
\setlength{\oddsidemargin}{-.8in}
\begin{figure}\flushleft
\begin{minipage}{7.5in}
\begin{center}
\end{center}
\caption{Plots (a) to (g) are the $\bar{X}-\bar{Y}$ probability densities
for the estimated density matrix in Simulation 2 at the times $\tau=4n\pi$
where $n=0,..,6$.}
\end{minipage}
\end{figure}
\newpage
\setlength{\oddsidemargin}{0in}
\begin{figure}
\begin{minipage}{6in}
\begin{center}
\epsfig{file=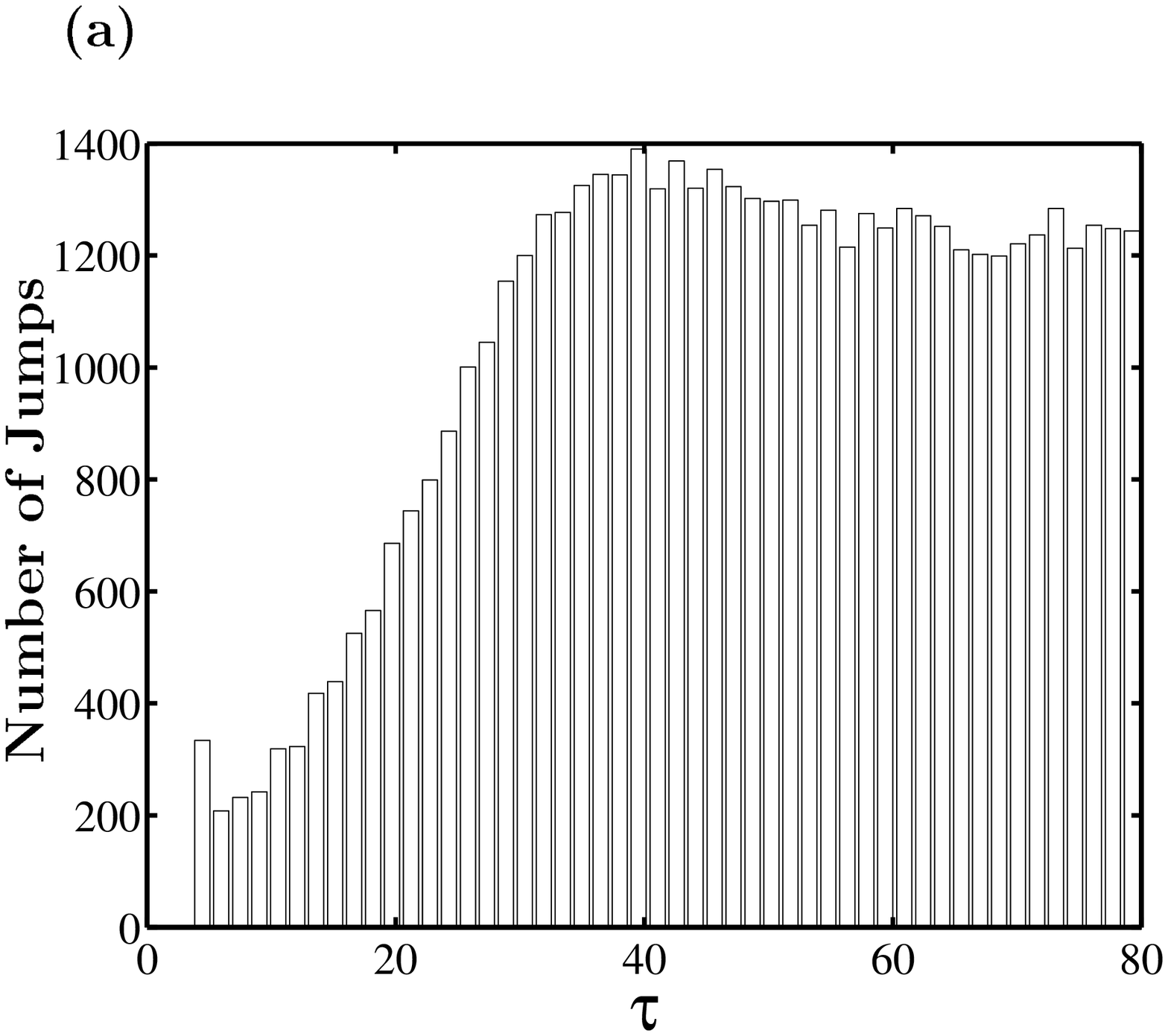,width=8cm}\\
\epsfig{file=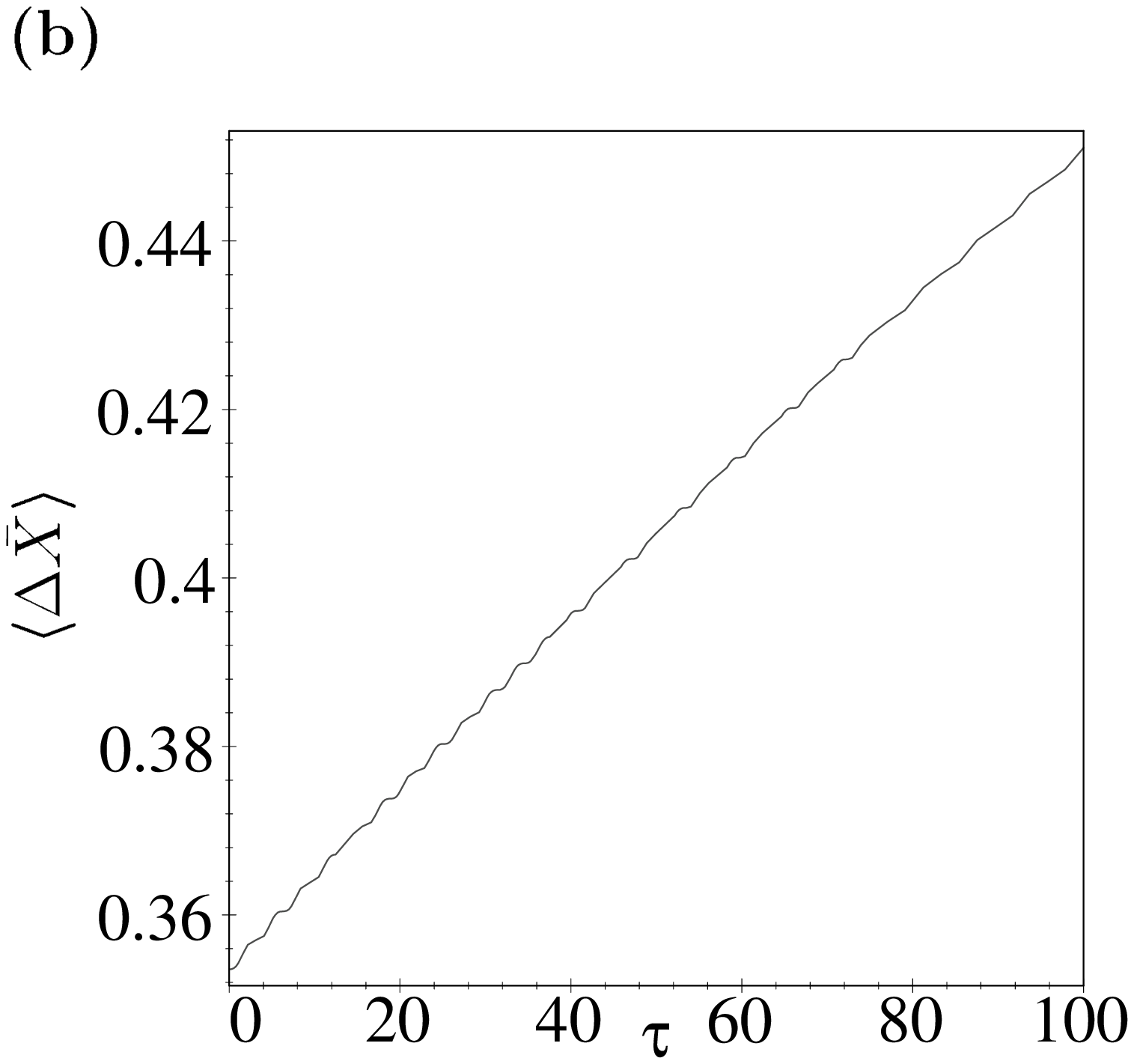,width=8cm}
\end{center}
\caption{Plot (a) is the histogram of jump frequencies for Simulation 2, i.e. numer of jumps which occurred between time $\tau$ and $\tau+5$,
while (b) plots $\langle \Delta \bar{X}(\tau )\rangle $ as an analytic
comparison to Simulation 1. Here we have neglected the cross terms and
recoil momentum kick in the master equation. This is solved analytically.
See Appendix D for details.}
\label{analytic}
\end{minipage}
\end{figure}


\begin{references}
\bibitem{INTEREST}  N. B. Simpson, L. Allen and M. J. Padgett, J. Mod. Opt.
{\bf 43}, 2485 (1996); V. E. Lembessis, L. Allen and M. Babiker,{\it ``Orbital angular
momentum effects on atoms: A density-matrix theory''}, in {\em Coherence and
Quantum Optics}, eds. Eberly, Mandel and Wolf, (Plenum, NY 1996).

\bibitem{FRIESE}  H. He, M. E. J. Friese, N. R. Heckenberg and H.
Rubinsztein-Dunlop, Phys. Rev. Lett. {\bf 75}, 826 (1995); M. E. J. Friese,
J. Engler, H. Rubinsztein-Dunlop and N. Heckenberg, Phys. Rev. A {\bf 54},
1593 (1996).

\bibitem{ERTMER_TRAP}  M. Schiffer, G. Wokurka, M. Rauner, S. Kuppens, T.
Slawinski, M. Zinner, K. Sengstock, W. Ertmer, {\it ``Holographically
designed light-fields as elements for Atom Optics''}, in IQEC '96 Technical
Digest, Sydney, Australia, (Optical Soc. America, ISBN 1-55752-459-9, 1996).

\bibitem{ALAN_INTRO}  M. Babiker, V. E. Lembessis, W. K. Lai,L. Allen, Opt.
Commun. {\bf 123}, 523 (1996).

\bibitem{JAPAN}  H. Ito, T. Nakata, K. Sakaki, W. Jhe and M. Ohtsu, {\it %
``Experiments on atom guidence with evansecent waves''}, in IQEC '96
Technical Digest, Sydney, Australia, (Optical Soc. America, ISBN
1-55752-459-9, 1996).

\bibitem{SPIRAL}  G. A. Turnbull, D. A. Robertson, G. M. Smith, L. Allen and
M. J. Padgett, Opt. Commun. {\bf 127}, 183 (1996).

\bibitem{CYLLENS}  C. Tamm and C. O. Weiss, J. Opt. Soc. Am. {\bf B7}, 1034
(1990); M. W. Beijersbergen, L. Allen, H. E. L. O. van der Veen and J. P.
Woerdman, Opt. Commun. {\bf 96}, 123 (1993).

\bibitem{RODNEY}  N. R. Heckenberg, R. McDuff, C. P. Smith and A. G. White,
Optics Lett. {\bf 17}, 221 (1992); N. R. Heckenberg, R. McDuff, C. P. Smith,
H. Rubinsztein-Dunlop and M. J. Wegener, Opt. and Quant. Elec. {\bf 24},
S951 (1992).

\bibitem{ENK}  S. J. van Enk and G. Nienhuis, Opt. Commun. {\bf 94}, 147
(1992).

\bibitem{POWER_JULY}  W. L. Power, L. Allen, M. Babiker and V. E. Lembessis,
Phys. Rev. A {\bf 52}, 479 (1995).

\bibitem{DYRTING}  S. Dyrting, Phys. Rev. A {\bf 53}, 2522 (1996).

\bibitem{KAZENTSEV}  A. P. Kazentsev, G. I. Surdutovich and V. P. Yakovlev, 
{\it Mechanical Action of Light on Atoms}, (World Scientific, Singapore,
1990).

\bibitem{RUSSIAN}  A. P. Kazantsev, V. S. Smirnov, A. M. Tumaikin and I. A.
Yagofarov, Opt. Spectrosc. {\bf 58}, 303 (1985).

\bibitem{ALLEN_MAY_1996}  L. Allen, V. E. Lembessis and M. Babiker, Phys.
Rev. A {\bf 53}, R2937 (1996).

\bibitem{VANENK_1994}  S. J. van Enk and G. Nienuis, J. Mod. Opts. {\bf 41},
963 (1994).

\bibitem{BETH}R. A. Beth, Phys. Rev. {\bf 50}, 115 (1936).

\bibitem{NEWINTEREST}N. B. Simpson, K. Dholakia, L. Allen, and M. J.
Padgett, Opt. Lett. {\bf 22}, 52, (1997).

\bibitem{QT}  J. Dalibard, Y. Castin and K. M\o lmer, Phys. Rev. Lett. {\bf %
68}, 580 (1992).

\bibitem{NUMERICAL_RECIPIES}  W.~H. Press, B.~P. Flannery, S.~A. Teukolsky
and W.~T. Vetterling, {\it Numerical Recipes}, (Cambridge University Press ,
Cambridge, 1990).

\bibitem{LENS}  M. Holland, S. Marksteiner, P. Marte and P. Zoller, Phys.
Rev. Lett. {\bf 76}, 3683 (1996).

\bibitem{FOURIER}  M. Wilkens, E. Schumacher and P. Meystre, Opt. Commun. 
{\bf 86}, 34 (1991).

\bibitem{WILKOX}  R. M. Wilkox, J. Math. Phys. {\bf 8}, 962 (1967).

\bibitem{HYPERGEO}  I. S. Gradshteyn and I. M. Ryzkih, {\it Tables of
Integrals, Series, and Products}, (Academic Press, London, 1980), section
9.100.

\bibitem{HALLIWELL_DOWKER}  H. F. Dowker and J. J. Halliwell, Phys. Rev. A 
{\bf 46}, 1580 (1992).

\bibitem{AZIMUTHAL}  L. Allen, M. Babiker and W. Power, Opt. Commun. {\bf 112%
}, 141 (1994).
\end{references}
\end{document}